\documentclass{article}

\usepackage{arxiv}
\usepackage{hyperref}       
\usepackage[utf8]{inputenc} 
\usepackage[T1]{fontenc}    
\usepackage{url}            
\usepackage{booktabs}       
\usepackage{nicefrac}       
\usepackage{microtype}      
\usepackage{lipsum}         
\usepackage{graphicx}
\usepackage{doi}
\usepackage{comment}
\usepackage[square,sort,comma,numbers]{natbib}
\usepackage{amsmath,amssymb,amsfonts}
\usepackage{cleveref}       
\usepackage[table]{xcolor}
\usepackage{xcolor}
\newcommand\crule[3][black]{\textcolor{#1}{\rule{#2}{#3}}}

\title{Energy-based PINNs for solving coupled field problems: concepts and application to the multi-objective optimal design of an induction heater}

\newif\ifuniqueAffiliation
\uniqueAffiliationtrue

\ifuniqueAffiliation 
\author{ \href{https://orcid.org/0000-0002-5803-3150}{\includegraphics[scale=0.06]{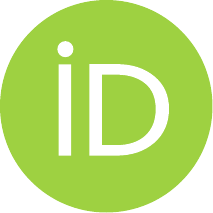}\hspace{1mm}Marco ~Baldan} \\
	Department of Optimization \\
	Fraunhofer ITWM\\
	Kaiserslautern, Germany \\
	\texttt{marco.baldan@itwm.fraunhofer.de} \\
	\And
	\href{https://orcid.org/0000-0001-5293-1809}{\includegraphics[scale=0.06]{orcid.pdf}\hspace{1mm}Paolo Di~Barba} \\
	Department of Electrical, Computer \& Biomedical Engineering \\
	University of Pavia\\
	Pavia, Italy \\
	\texttt{paolo.dibarba@unipv.it} \\
}
\else
\usepackage{authblk}

\newbox{\orcid}\sbox{\orcid}{\includegraphics[scale=0.06]{orcid.pdf}} 
\author[1]{%
	\href{https://orcid.org/0000-0000-0000-0000}{\usebox{\orcid}\hspace{1mm}David S.~Hippocampus\thanks{\texttt{hippo@cs.cranberry-lemon.edu}}}%
}
\author[1,2]{%
	\href{https://orcid.org/0000-0000-0000-0000}{\usebox{\orcid}\hspace{1mm}Elias D.~Striatum\thanks{\texttt{stariate@ee.mount-sheikh.edu}}}%
}
\affil[1]{Department of Computer Science, Cranberry-Lemon University, Pittsburgh, PA 15213}
\affil[2]{Department of Electrical Engineering, Mount-Sheikh University, Santa Narimana, Levand}
\fi

\renewcommand{\shorttitle}{\textit{arXiv} Template}

\hypersetup{
pdftitle={A template for the arxiv style},
pdfsubject={q-bio.NC, q-bio.QM},
pdfauthor={David S.~Hippocampus, Elias D.~Striatum},
pdfkeywords={First keyword, Second keyword, More},
}

\begin{document}
\maketitle

\begin{abstract}
	Physics-informed neural networks (PINNs) are neural networks (NNs) that directly encode model equations, like Partial Differential Equations (PDEs), in the network itself. While most of the PINN algorithms in the literature minimize the local residual of the governing equations, there are energy-based approaches that take a different path by minimizing the variational energy of the model. We show that in the case of the steady thermal equation weakly coupled to magnetic equation, the energy-based approach displays multiple advantages compared to the standard residual-based PINN: it is more computationally efficient, it requires a lower order of derivatives to compute, and it involves less hyperparameters. The analyzed benchmark problems are the single- and multi-objective optimal design of an inductor for the controlled heating of a graphite plate. The optimized device is designed involving a multi-physics problem: a time-harmonic magnetic problem and a steady thermal problem. For the former, a deep neural network solving the direct problem is supervisedly trained on Finite Element Analysis (FEA) data. In turn, the solution of the latter relies on a hypernetwork that takes as input the inductor geometry parameters and outputs the model weights of an energy-based PINN (or ePINN). Eventually, the ePINN predicts the temperature field within the graphite plate.
\end{abstract}


\section{Introduction}\label{sec1}
Physics-Informed Neural Networks are neural networks that encode model equations, like PDEs, as a component of the neural network itself \cite{r1}. PINNs are nowadays used to solve PDEs, stochastic PDEs, fractional, and integral-differential equations in a wide range of applications in both science and engineering. For the most part of the available literature, PINNs are trained by minimizing a neural network loss function that includes terms reflecting the initial (ICs) and boundary conditions (BCs) along the space-time domain's boundary and the PDE residual at selected points in the domain (called collocation points) \cite{r2}. However, some researchers adopt the variational form of PDEs and minimize the corresponding energy functional \cite{r3}. This approach has been successfully adopted in, e.g., computational mechanics \cite{r4}, \cite{r5}, \cite{r6}, and it could be applied in case the governing PDEs can be derived from a known functional. This is the case, among others, for the Laplace, Poisson and Helmholtz equations \cite{r7}.

Induction heating is generally used for thermal processes of materials \cite{r8}. In fact it allows a good temperature control of the work-piece temperature with a high efficiency. In this field, the solution of weakly coupled (quasi-static) magnetic and thermal problems is mandatory \cite{r9}. The magnetic problem is typically solved in the frequency domain, while, depending on the application, the thermal solution is either time transient or steady-state. In the present work, we will focus on the steady case. There is little literature available involving PINNs for solving quasi-static time-harmonic magnetic problems \cite{r10}. On the contrary, albeit not involving induction heating applications, PINNs have been often adopted for the solution of time transient \cite{r11}, \cite{r12} and steady \cite{r13}, \cite{r14} heat conduction problems. However, in the existing literature, PINN's training for the steady heat problem relies on the minimization of the PDE residual \cite{r13}, \cite{r14}. In this work, on the one hand, we will show the benefits of adopting the energy-based approach compared to the strong residual method. First, it is more computationally efficient, i.e., with the same network architecture and the same amount of collocation points, less training epochs are required to achieve a prescribed accuracy in the temperature field. Second, it involves a lower order of derivatives (first rather than second), therefore allowing material discontinuities in the domain. The first and second advantages imply a significant
speed up in network training. Finally, it requires less (or no) hyperparameters in the loss, making the training more robust. On the other hand, we want to assess new approaches in the induction heating framework. The current work differs from already existing contributions \cite{r10}, \cite{r15}, \cite{r16}, \cite{r17}. In \cite{r10}, authors introduce PINNs in a magnetic-thermal coupled problem, but still relying on the conventional residual-based approach and considering a toy one-dimensional example. In \cite{r15}, a convolutional neural network (CNN) is trained on FEA generated temperature maps to solve an identification problem. \cite{r16} deals with an orthogonal decomposition method to build a data-driven digital twin. Finally, in \cite{r17} it is investigated
the explainability of black-box models trained on real-world data concerning surface hardening. In this work, the analyzed benchmark problem is the design of an inductor for the controlled heating of a graphite plate (adapted from \cite{r18}). A first neural network (called mNN) approximates the time-harmonic magnetic solution and it maps the design variables (or system parameters, in the current benchmark it is the inductor geometry) to the Joule loss distribution. mNN is supervisely trained on FEA simulations. Second, an energy-based PINN is trained (tePINN) to solve the temperature field for a particular set of system parameters. Adopting the residual-based approach allows to reduce of one order of magnitude the training cost. Finally, a hypernetwork \cite{r19} (called tHNN), which takes as input the design variables and outputs the model weights of a tePINN \cite{r20}, is used. In turn, the tePINN predicts the temperature field within the graphite plate. It is worth underlying that both tePINN and tHNN do not rely on any (e.g., FEA) data. Moreover, in order to train the tHNN, mNN is needed to provide the joule loss distribution, that is the source term in the heat equation. It follows that, once trained, the tHNN provides the fast (i.e., it only requires two forward passes) solution of a parametric coupled problem. By combining the tHNN with an optimizer, it gives the opportunity of quickly solving an inverse (e.g., optimal design) problem.  We compare the accuracy and the computational burden between the proposed approach and more traditional methods (e.g., FEA-based) in the case of a single and a multi-objective optimal design problem. The present work is both accurate and computationally original when solving numerous single- and multi-objective design problems (i.e., involving different objectives) quickly. Moreover, due to the availability of exact derivatives, it is convenient for constrained design tasks.  \\
The paper is structured as follows: section \ref{sec2} introduces the energy-based physics-informed neural network framework. Section \ref{sec3} describes the benchmark problem and the use of neural networks to solve single- and multi-objective design problems. Finally, section \ref{sec4} concludes the paper.

\section{Energy-based Physics-Informed Neural Networks}\label{sec2}

Consider, without loss of generality, the 2D boundary value problem:

\begin{equation}
	-\frac{\partial}{\partial x}\left(\alpha_{x} \frac{\partial \phi}{\partial x}\right)-\frac{\partial}{\partial y}\left(\alpha_{y} \frac{\partial \phi}{\partial y}\right) + \beta \phi = f \text { in } \Omega
	\label{eq1}
\end{equation}

where $\phi$ is the unknown function, $\alpha_{x}$, $\alpha_{y}$, and $\beta$ are known
parameters, which could be position-dependent functions, associated
with the physical properties of the domain, and $f$ is the source or excitation function \cite{r7}. We assume that $\alpha_{x}$, $\alpha_y$ are continuous over the entire domain. The boundary
conditions to be considered are:

\begin{equation}
	\phi = p \; \; \text{on} \; \Gamma_1 
	\label{eq2}
\end{equation}

and 

\begin{equation}
	\left(\alpha_x \frac{\partial \phi}{\partial x} \vec{x}+\alpha_y \frac{\partial \phi}{\partial y} \vec{y}\right) \cdot \vec{n}+\gamma \phi=q \; \text { on } \Gamma_2
	\label{eq3}
\end{equation}

where $\Gamma = \Gamma_1 + \Gamma_2$ denotes the contour enclosing the area, $\vec{n}$ is the normal unit vector, $\gamma$, $p$ and $q$ are known parameters associated with physical properties on the boundary. The variational problem equivalent to the boundary-value problem
is given by \cite{r7}:

\begin{equation}
	\left\{\begin{array}{l}
		\delta F(\phi)=0 \\
		\phi=p \text { on } \Gamma_1
	\end{array}\right.
	\label{eq4}
\end{equation}

with functional $F$ defined as:

\begin{equation}
	\begin{array}{r}
		F(\phi)=\frac{1}{2} \iint_{\Omega}\left[\alpha_x\left(\frac{\partial \phi}{\partial x}\right)^2+\alpha_y\left(\frac{\partial \phi}{\partial y}\right)^2+\beta \phi^2\right] d \Omega+ 
		\int_{\Gamma_2}\left(\frac{\gamma}{2} \phi^2-q \phi\right) d \Gamma-\iint_{\Omega} f \phi \, d \Omega .
	\end{array}
	\label{eq5}
\end{equation}

It is assumed all parameters and functions to be real-valued, therefore this functional is real-valued too. It can be shown that Eq. \ref{eq4} has a unique solution, i.e., the functional has a unique minimum. In case of the 2D steady heat conduction equation, $\phi$ corresponds to the temperature $T$, $f$ denotes the heat source density $Q$. In a weakly coupled electro-thermal problem the source term $Q$ takes induced current loss into account (more details in section \ref{sec3c}). Considering an anisotropic material $\alpha_{x} = \alpha_y = \lambda$ ($\lambda$ is the heat conductivity) subject to Dirichlet and Robin BCs ($p = T_{ref}$ is a specified temperature, having  $\gamma=h$, where $h$
is the overall heat transfer coefficient, $q = h T_0$, $T_0$ being the external temperature), the system of governing equations becomes:

\begin{equation}
	-\frac{\partial}{\partial x}\left(\lambda \frac{\partial T}{\partial x}\right)-\frac{\partial}{\partial y}\left(\lambda \frac{\partial T}{\partial y}\right)=Q \text { in } \Omega
	\label{eq6}
\end{equation}

with 

\begin{equation}
	T = T_{ref} \text { in } \Gamma_1
	\label{eq7}
\end{equation}

and

\begin{equation}
	\left(\lambda \frac{\partial T}{\partial x} \vec{x}+\lambda \frac{\partial T}{\partial y} \vec{y}\right) \cdot \vec{n}+h T=h T_0 \text { on } \Gamma_2
	\label{eq8}
\end{equation}

with the functional:

\begin{equation}
	\begin{array}{r}
		F(T)=\frac{1}{2} \iint_{\Omega}\left[\lambda\left(\frac{\partial T}{\partial x}\right)^2+\lambda\left(\frac{\partial T}{\partial y}\right)^2\right] d \Omega+ 
		\int_{\Gamma_2}\left(\frac{h}{2} T^2-h T_0 T\right) d \Gamma-\iint_{\Omega} Q T d \Omega .
	\end{array}
	\label{eq9}
\end{equation}

Introducing now the energy-based PINN $\mathcal{N}_p$ (tePINN) with weights (including biases) $\boldsymbol{\theta}_{\boldsymbol{p}}$, its loss function to minimize corresponds to the functional:

\begin{equation}
	\begin{array}{r}
		L_p\left(\boldsymbol{\theta}_{\boldsymbol{p}}\right)=\frac{1}{2} \iint_{\Omega}\left[\lambda\left(\frac{\partial \hat{T}\left(\mathrm{x}, \boldsymbol{\theta}_{\boldsymbol{p}}\right)}{\partial x}\right)^2+\lambda\left(\frac{\partial \hat{T}\left(\mathrm{x}, \boldsymbol{\theta}_{\boldsymbol{p}}\right)}{\partial y}\right)^2\right] d \Omega 
		+\int_{\Gamma_2}\left(\frac{h}{2} \hat{T}^2\left(\mathbf{x}, \boldsymbol{\theta}_{\boldsymbol{p}}\right)-h T_0 \hat{T}\left(\mathbf{x}, \boldsymbol{\theta}_{\boldsymbol{p}}\right)\right) d \Gamma + \\
		-\iint_{\Omega} Q \hat{T}\left(\mathbf{x}, \boldsymbol{\theta}_{\boldsymbol{p}}\right) d \Omega
	\end{array}
	\label{eq10}
\end{equation}

Derivatives are obtained via automatic differentiation (AD). The predicted temperature field is:

\begin{equation}
	\hat{T}(\mathbf{x}; \boldsymbol{\theta}_{\boldsymbol{p}}) = a+b \mathcal{N}_p(\mathbf{x}; \boldsymbol{\theta}_{\boldsymbol{p}})
	\label{eq10a}
\end{equation}

with $a$ and $b$ user-defined positive scalar values. $a$ is an offset and it corresponds to a reference average temperature, $b$ is a scaling factor. Either trapezoidal rule or Monte Carlo integration can be used for computing the integral by sampling the domain with either randomly or uniformly spaced points. However, with this approach, a large number of integration points are usually required to obtain accurate results. To address this issue, we prefer the Gauss quadrature rule, with a sampling strategy that is close to a meshing step in conventional FEA (a weakness in this case of the energy-based approach, especially in case of complex geometries):

\begin{equation}
	\begin{aligned}
		L_p^E\left(\boldsymbol{\theta}_{\boldsymbol{p}}\right)= & \frac{1}{2} \sum_{i=1}^{N_{\Omega}} A_i \sum_{k=1}^{N_G} \lambda_{i, k}\left[\left(\frac{\partial \hat{T}\left(\mathbf{x}_{i, k}, \boldsymbol{\theta}_{\boldsymbol{p}}\right)}{\partial x}\right)^2+\left(\frac{\partial \hat{T}\left(\mathbf{x}_{i, k}, \boldsymbol{\theta}_{\boldsymbol{p}}\right)}{\partial y}\right)^2\right]+ \\
		+ & \sum_{i=1}^{N_{\Gamma_2}} l_i \sum_{k=1}^{N_g}\left(\frac{h}{2} \hat{T}^2\left(\mathbf{x}_{i, k}, \boldsymbol{\theta}_{\boldsymbol{p}}\right)-h T_0 \hat{T}\left(\mathbf{x}_{i, k}, \boldsymbol{\theta}_{\boldsymbol{p}}\right)\right)
		 - \sum_{i=1}^{N_{\Omega}} A_i \sum_{k=1}^{N_G} Q_{i, k} \hat{T}\left(\mathbf{x}_{i, k}, \boldsymbol{\theta}_{\boldsymbol{p}}\right)
	\end{aligned}
	\label{eq11}
\end{equation}

where $N_{\Omega}$ and $N_{\Gamma_2}$ is the number of discretization elements of the domain and the $\Gamma_2$ boundary, respectively. $N_G$ denotes the number of integration points for each element, $A$ and $l$ are area and length of the corresponding element, respectively. The discretization (i.e., sampling points) remains unchanged while training. For the sake of completeness, we mention that the residual-based PINN loss would be:

\begin{equation}
	\begin{gathered}
		L_{p}^{R}\left(\boldsymbol{\theta}_{\boldsymbol{p}}\right)=\frac{1}{S_{\Omega}} \sum_{j=1}^{S_{\Omega}}\left[-\frac{\partial}{\partial x}\left(\lambda_j \frac{\partial \hat{T}\left(\mathbf{x}_j, \boldsymbol{\theta}_{\boldsymbol{p}}\right)}{\partial x}\right)\right. + 
		\left.-\frac{\partial}{\partial y}\left(\lambda_j \frac{\partial \hat{T}\left(\mathbf{x}_j, \boldsymbol{\theta}_{\boldsymbol{p}}\right)}{\partial y}\right)-Q_j\right]^2+\frac{\eta_{1}}{S_{\Gamma_1}} \sum_{j=1}^{S_{\Gamma_1}}\left[\hat{T}\left(\mathbf{x}_j, \boldsymbol{\theta}_{\boldsymbol{p}}\right)-T_{r e f}\right]^2 \\
		+ \frac{\eta_{2}}{S_{\Gamma_2}} \sum_{j=1}^{S_{\Gamma_2}}\left[\lambda_j \frac{\partial \hat{T}\left(\mathbf{x}_j, \boldsymbol{\theta}_{\boldsymbol{p}}\right)}{\partial n}+h \hat{T}\left(\mathbf{x}_j, \boldsymbol{\theta}_{\boldsymbol{p}}\right)-h T_0\right]^2
	\end{gathered}
	\label{eq12}
\end{equation}

with $S_{\Omega}$, $S_{\Gamma_1}$, $S_{\Gamma_2}$ the amount of collocation points within
the domain, on the $\Gamma_1$ and $\Gamma_2$ boundaries, respectively. It stands out that the residual-based loss involves second derivatives (instead first derivatives only appear in Eq. 11) and needs hyperparameters ($\eta_{1}$, $\eta_{2}$ - no one appears in the energy-based loss). The presented physics-informed neural network does not take any system parameter (like geometry or material properties) as input, therefore, when embedded in inverse problems, solving the PDE governing equations, it would require additional training for each new system parameter set. In this sense, transfer learning (TL) would help \cite{r23}. However, we prefer a hypernetwork (tHNN - "thermal" HNN) $\mathcal{N}_{h}$ \cite{r19} (whose weights are $\boldsymbol{\theta}_{\boldsymbol{h}}$) that receives system parameters ($\xi$) and outputs the network weights of the tePINN ("thermal" ePINN), which in turn provides the solution of the direct thermal problem \cite{r20}, \cite{r21}. In fact, tePINN predicts the temperature evaluated at $\textbf{x}$. The tHNN loss function is:

\begin{equation}
	L_h\left(\boldsymbol{\theta}_{\boldsymbol{h}}\right)=\sum_{i=1}^{N_{\xi}} L_p^E\left(\mathcal{N}_h\left(\boldsymbol{\xi}^i ; \boldsymbol{\theta}_{\boldsymbol{h}}\right)\right)
	\label{eq13}
\end{equation}

$N_{\xi}$ indicates the number of system parameter sets involved in a batch. The forward pass in the training of tHNN works as follows: for each system parameter set  $\boldsymbol{\xi}$ (i.e., inductor geometry), the tHNN outputs the weights and biases $\boldsymbol{\theta_{p}}$ of a tePINN. In turn, tePINN is supposed to provide the entire temperature field, evaluated at both locations within $\Omega$ and along $\Gamma_{1,2}$, for that parameter set. Assuming to know the heat source map $Q$ within the graphite plate (a "magnetic" network is adopted in this work for such a purpose - see \ref{sec3c}), the training loss could be estimated (Eq. \ref{eq11}). The forward pass is repeated with all the inductor geometries in the batch (the batch size is $N_{\xi}$) to obtain the tHNN loss function (Eq. \ref{eq13}). Gradient information (derivatives w.r.t. network weights and biases $\boldsymbol{\theta_{h}}$) are provided to the optimizer by AD.  More about the training in section \ref{sec3c}. Once trained, the hypernetwork is able to predict the temperature map for a given system parameter set. Therefore, solving the inverse problems with $\boldsymbol{\xi}$ as design variables is associated with a little computational cost. Moreover, it is straightforward to consider different new multiple objectives and/or constraints. Finally, the optimizer can take advantage of exact derivatives due to AD \cite{r21}. More details in section \ref{sec3c}.

\section{The benchmark problem}\label{sec3}

\subsection{Induction heating of a graphite plate}\label{sec3a}

The 2D benchmark (adapted from \cite{r18} - planar unlike axisymmetric geometry; this represents no limitation regarding the applications of the presented approach) includes a graphite (electrical and thermal properties at 1200$^{\circ}$C: $\rho_{g}$ = 7.76 10$^{-6}$ $\Omega$m, $\mu_{r,g}=1$, $\lambda$ = 60 W/(mK)) plate of width 240 mm and thickness 14 mm (for symmetry reasons only half of it is considered) and a 8-turn induction heater (made of copper, $\rho_{c}$ = 2 10$^{-8}$ $\Omega$m, $\mu_{r,c}=1$) (Fig. \ref{fig1}). All turns, series connected, carry a current of 400 Arms at 4250 Hz. The plate is supposed to achieve a steady-state average temperature around 1100-1150$^{\circ}$C, as required by the industrial process. Due to the desired temperature uniformity, material parameters are not temperature dependent \cite{r18}.

\subsection{Field model: a coupled magnetic-thermal analysis}\label{sec3b}
The direct problem involves a time-harmonic quasi-static magnetic problem (to evaluate the power density in the graphite plate) coupled to a steady-state thermal problem (to evaluate the temperature profile). The magnetic problem is formulated in terms of the phasor of the magnetic vector potential $\dot{\textbf{A}}$):

\begin{equation}
	\nabla^{2} \dot{\textbf{A}} - j\omega \mu \rho^{-1} \dot{\textbf{A}} = - \mu \dot{\textbf{J}}
	\label{eq14}
\end{equation}

where $\dot{\textbf{J}}$ is the phasors of the current density, $\mu$ is the magnetic permeability, $\rho$ is the electrical resistivity and $\omega$ is the magnetic field pulsation. The solutions of the magnetic and thermal problem are weakly coupled by means of the source
term:

\begin{equation}
	Q = \rho^{-1} \omega^2 |\dot{\textbf{A}}|^2
	\label{eq15}
\end{equation}

of the thermal equation (Eqs. \ref{eq6}-\ref{eq8}). The thermal domain is the graphite plate. Along the axis of symmetry (i.e., at x=0), Neumann BCs are imposed. At the other bounds, Robin BCs are adopted with overall heat exchange coefficient of 50 W/(m$^2$K) and external temperature of 50$^{\circ}$C (Fig. \ref{fig1}). The overall heat exchange coefficient is meant to already include the radiation losses.

\subsection{Approximating and solving the direct problem}\label{sec3c}
\subsubsection{Magnetic analysis}\label{sec3c1}

\begin{figure}[!t]
	\centering{\includegraphics[width=0.7\columnwidth]{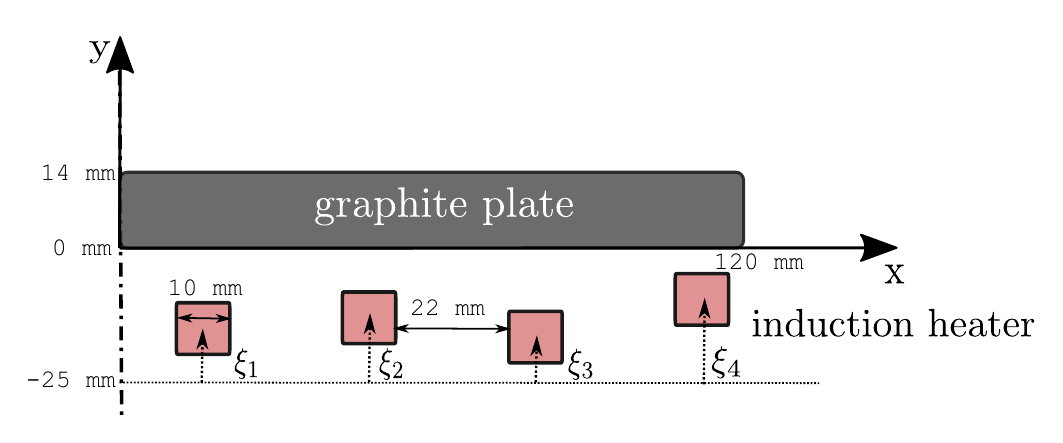}}
	\caption{ Benchmark setup (half of it due to symmetry).\label{fig1}}
\end{figure}

\begin{figure}[!t]
	\centering{\includegraphics[width=0.5\columnwidth]{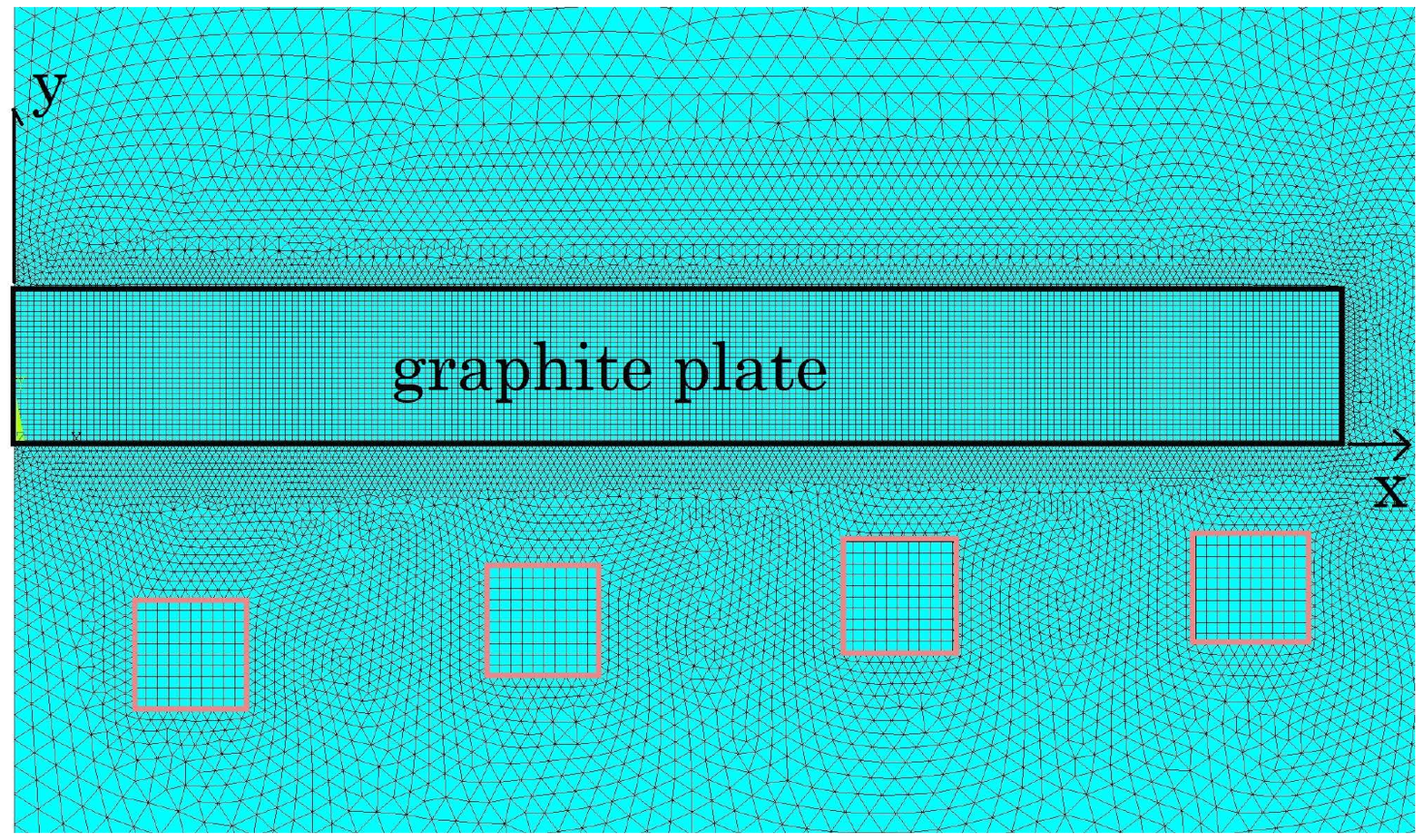}}
	\caption{Particular of the mesh (graphite plate and induction heater) used in the FEA.\label{fig1a}}
\end{figure}

A fully-connected neural network (mNN - "magnetic" NN) with 3 hidden layers of 128 neurons each is supervisedly trained to map the inductor geometry (i.e., the y-position of each turn - $\pmb{\xi} = [\xi_1, \xi_2, \xi_3, \xi_4]$ - Fig. \ref{fig1}) and the space position (x, y) with the heat source density ($Q$) (Eq. \ref{eq15}) within the graphite plate. The neural architecture search relies on a trial and error approach: two other architectures with 3 hidden layers of 64 and 256 neurons, respectively, are trained. FEA simulations run at different system parameter sets, with a uniform discretization (range 5 to 15 mm) covering 6 values for each variable and a total of 1296 magnetic solutions, constitutes the training data-set. Since the elements of the mesh in the FEA (Fig. \ref{fig1a}) do not coincide with those adopted for tePINN's or tHNN's (Eq. \ref{eq11}) training, each FEA solution is projected to the Gauss integration points adopting a piecewise cubic, continuously differentiable, and approximately curvature-minimizing polynomial approximation (see scipy.interpolate.griddata \cite{r22}). Interpolated data actually serves as training data-set for the mNN. 30 additional random inductor geometries and their corresponding FEA simulations form the test-data set, instead. FEA is performed with a commercial software \cite{r24}. The model has $5 \, 10^4$ linear elements (Fig. \ref{fig1a}). 

\begin{figure}[!t]
	\centering{\includegraphics[width=0.6\columnwidth]{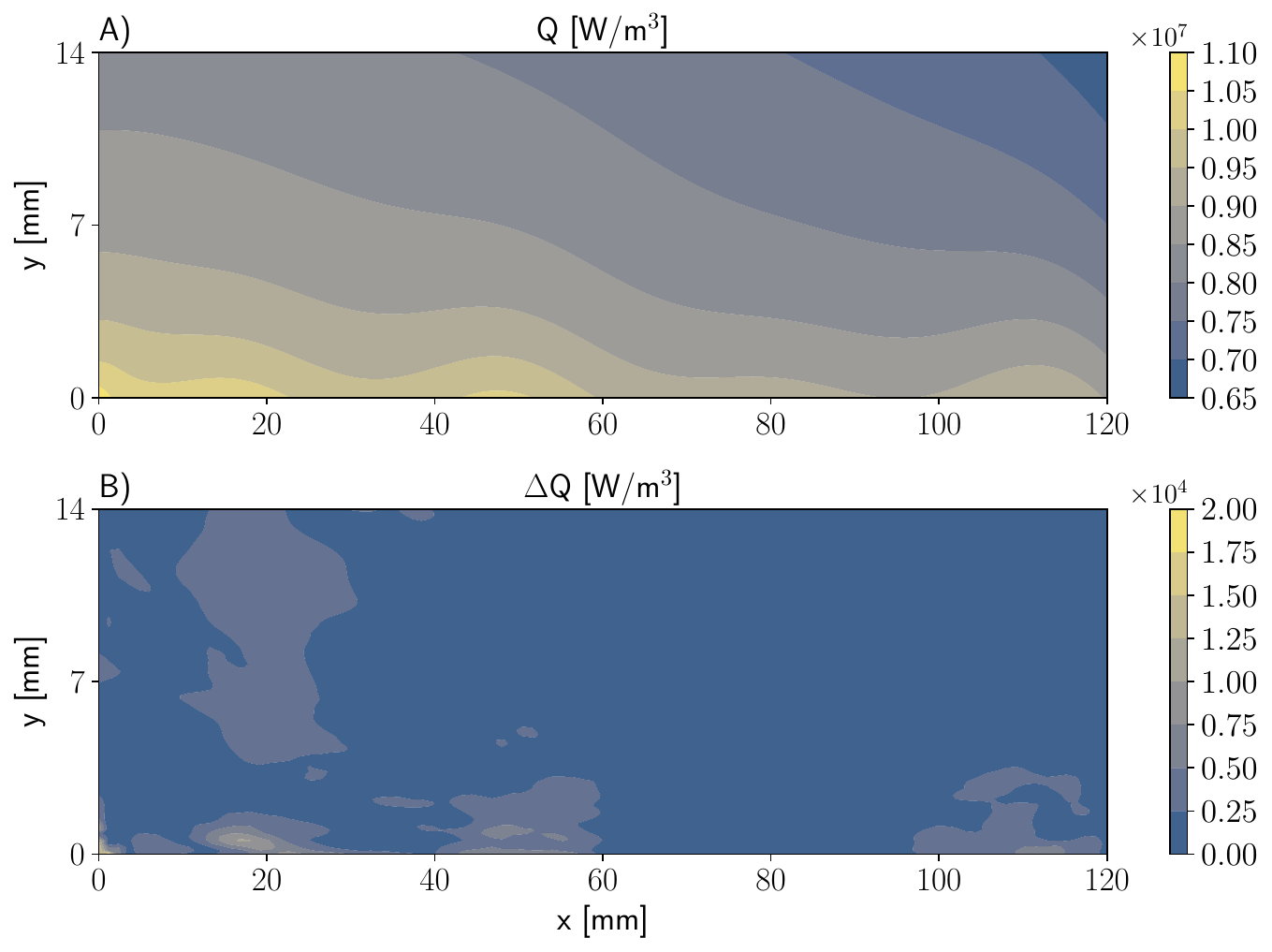}}
	\caption{Estimating the accuracy of the mHH for a random test inductor geometry ($\xi_1$ = 13.63 mm, $\xi_2$ = 13.59 mm,  $\xi_3$ = 9.59 mm, $\xi_4$ = 12.67 mm).
		A) Joule losses within the graphite plate. B) Absolute error taking the FEA solution as ground truth.\label{fig1b}}
\end{figure}

The activation function of mNN is $\max(0; x^3)$. It is trained on $4 \, 10^4$ epochs using the ADAM optimizer. All machine learning computations are carried out on TensorFlow \cite{r25}. The mean absolute error (MAE) on the test-set (evaluated at 10$^4$ positions of each of the 30 random inductor geometries) is equal to $1.44 \; 10^3$ W/m$^3$, which means 0.017\% in percentage terms. The maximum absolute error is $3.88 \; 10^4$ W/m$^3$ (0.46\%) instead. The architecture with 128 neurons in each layer shows the best performance in the test-set and it will be used in the next analyses (Tab. \ref{tab0}). An exemplary Joule loss distribution is depicted in Fig.\ref{fig1b}.

\begin{table}
	\caption{Architecture search for mNN: average ($\overline{\Delta Q}$) and maximum (${\Delta Q}_{\max}$) absolute prediction error on the test-set \label{tab0}}
	\centering
	\begin{tabular}{llll}
		\toprule
		Hidden  & Neurons  & $\overline{\Delta Q}$ & ${\Delta Q}_{\max}$ \\
		layers  & each layer & [W/m$^3$] & [W/m$^3$]  \\
		\midrule
		3  & 64  & $1.89 \; 10^3$ & $5.32 \; 10^4$   \\
		&      &  0.018\% & 0.63\%                 \\
		3  & 128  & $\mathbf{1.44 \; 10^3}$ &  $\mathbf{3.88 \; 10^4}$  \\
		&      &  $\mathbf{0.017\%}$ & $\mathbf{0.46\%}$     			\\
		3  & 256  & $1.93 \; 10^3$ & $4.34 \; 10^4$ \\
		&      &  0.023\% & 0.51\%     \\
		\bottomrule
	\end{tabular}
\end{table}

\subsubsection{Single thermal analysis - tePINN}\label{sec3c2}
\begin{figure}[!t]
	\centering{\includegraphics[width=0.6\columnwidth]{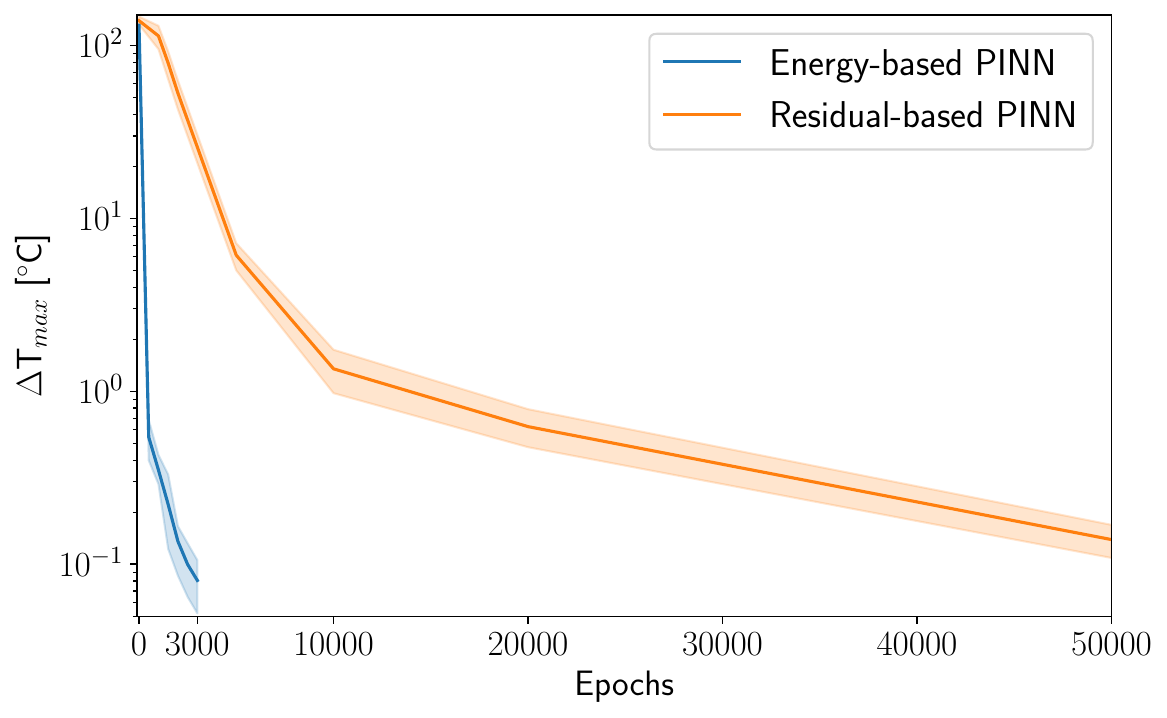}}
	\caption{Max absolute error in the temperature field solution (ground truth is the FEA solution). The residual-based PINN takes $\eta_{2}=10^3$ in the loss function \label{fig2}}
\end{figure}

The architecture of tePINN is a fully-connected neural network with 2 hidden layers with 64 neurons each. This architecture is the result of a trial and error approach (Tab. \ref{tab05}). Offset and scaling factors $a$ and $b$ (Eq. \ref{eq10a}) are set to 900$^{\circ}$C and 300$^\circ$C, respectively. We consider a specific inductor geometry ($\xi_1$ = 14.8 mm, $\xi_2$ = $\xi_3$ = $\xi_4$ = 15 mm) and use, only here, the FEA solution for the distribution of the heat source density (in particular, a cubic interpolation to the collocation and integration points - see scipy.interpolate.griddata \cite{r22}). In the thermal domain (i.e., the graphite plate), 10$^4$ elements (all equal) are considered. Integration relies on the 2-point Gaussian quadrature. For sake of comparison, a residual-driven PINN, having the same architecture, adopting the same amount of collocation points, and taking the same weight initialization, is trained as well. Considering the FEA solution as the ground truth, the maximum absolute error behaves in the two approaches much differently (Fig. \ref{fig2}).

\begin{figure}[h!]
	\centering{\includegraphics[width=0.6\columnwidth]{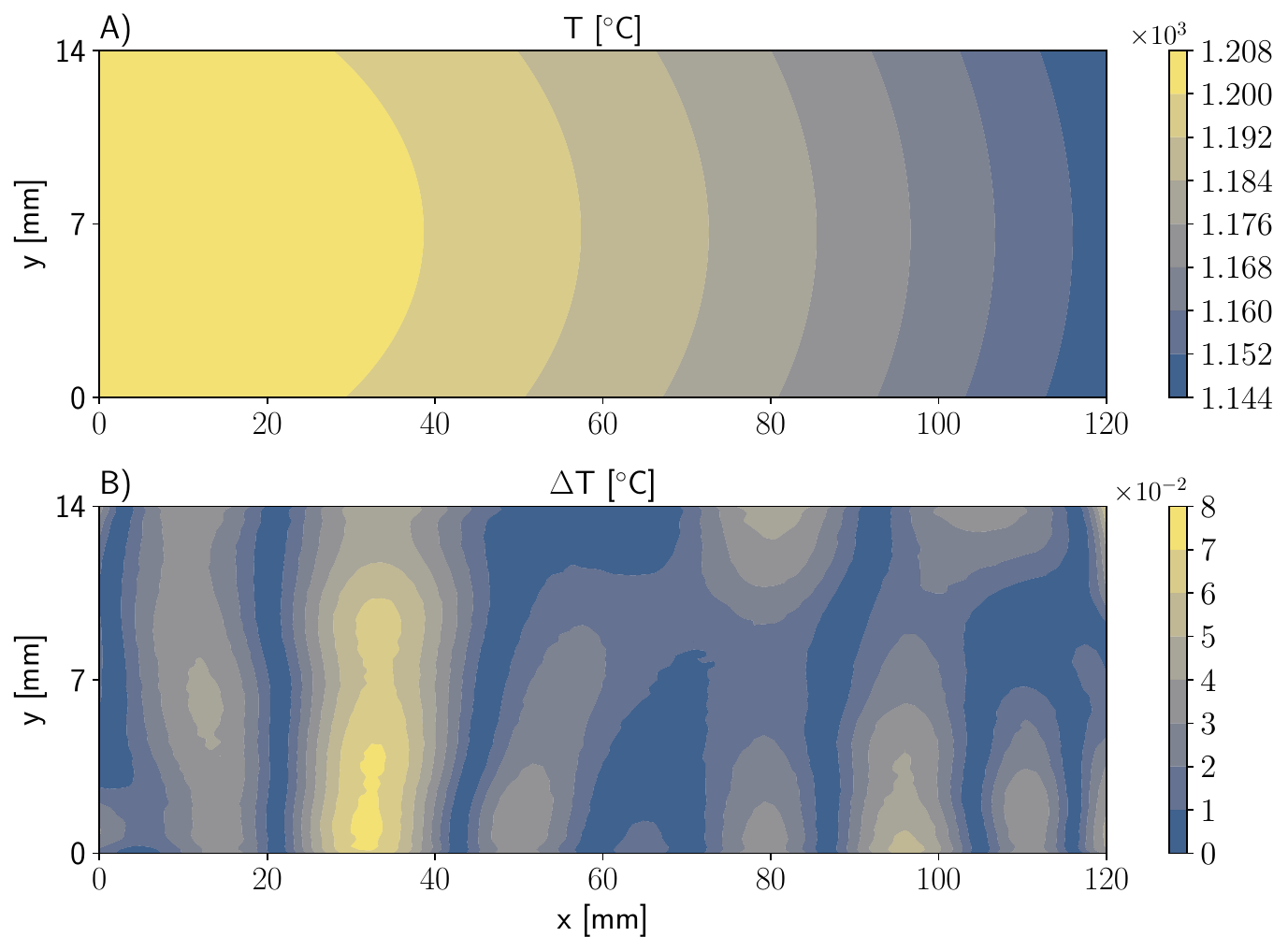}}
	\caption{Estimating accuracy of the tePINN prediction in case of a given inductor geometry ($\xi_1$ = 14.8 mm, $\xi_2$ = $\xi_3$ = $\xi_4$ = 15 mm). The heat source distribution comes from FEA solution. \label{fig3}
		A) Temperature field within the graphite plate. B) Absolute error taking the FEA solution as ground truth.}
\end{figure}

The energy-based approach shows a much faster learning and it characterized by a solution error around 0.1$^{\circ}$C after just 3 10$^3$ epochs. On the contrary, the residual-based approach takes approximately 5 10$^4$ epochs to achieve a comparable accuracy. Looking at the computation time on a personal laptop, the advantage of adopting the energy-based PINN is even larger (here approximately a factor 50), due to the lower order of derivatives involved. Since the residual-based loss (Eq. \ref{eq12}) involves hyperparameters (in this benchmark only $\eta_{2}$, since no Dirichlet BCs are imposed), obtained results concern the best achieved among $\eta_2$ = [10$^1$, 10$^2$, 10$^3$, 10$^4$, 10$^5$], namely with $\eta_{2}=10^3$. The temperature field and the absolute error are depicted in Fig. \ref{fig3}.

\begin{table}
	\caption{Architecture search for tePINN: (average on 5 training) maximum (${\Delta T}_{\max}$) absolute prediction error on given geometry \label{tab05}}
	\centering
	\begin{tabular}{llll}
		\toprule
		Hidden  & Neurons  & Network & ${\Delta T}_{\max}$  \\
		layers  & each layer & weights & [$^\circ$C]  \\
		\midrule
		2  & 16  & 337  & 0.359  \\
		&      &   & 0.033\%              \\
		2  & 24  & 697 & 0.186 \\
		&      &  &     0.017\% 			\\
		2  & 32  & 1185 & 0.134 \\
		&      &  & 0.012\%  \\
		2  & 64  & 4417 & $\mathbf{0.112}$ \\
		&      &   &  $\mathbf{0.010\%}$ 	\\
		2  & 128  & 17025  & 0.137 \\
		&      &   & 0.012\% \\
		\bottomrule
	\end{tabular}
\end{table}

\subsubsection{Parametric thermal analysis - tHNN}\label{sec3c3}
\begin{figure}[!t]
	\centering{\includegraphics[width=0.5\columnwidth]{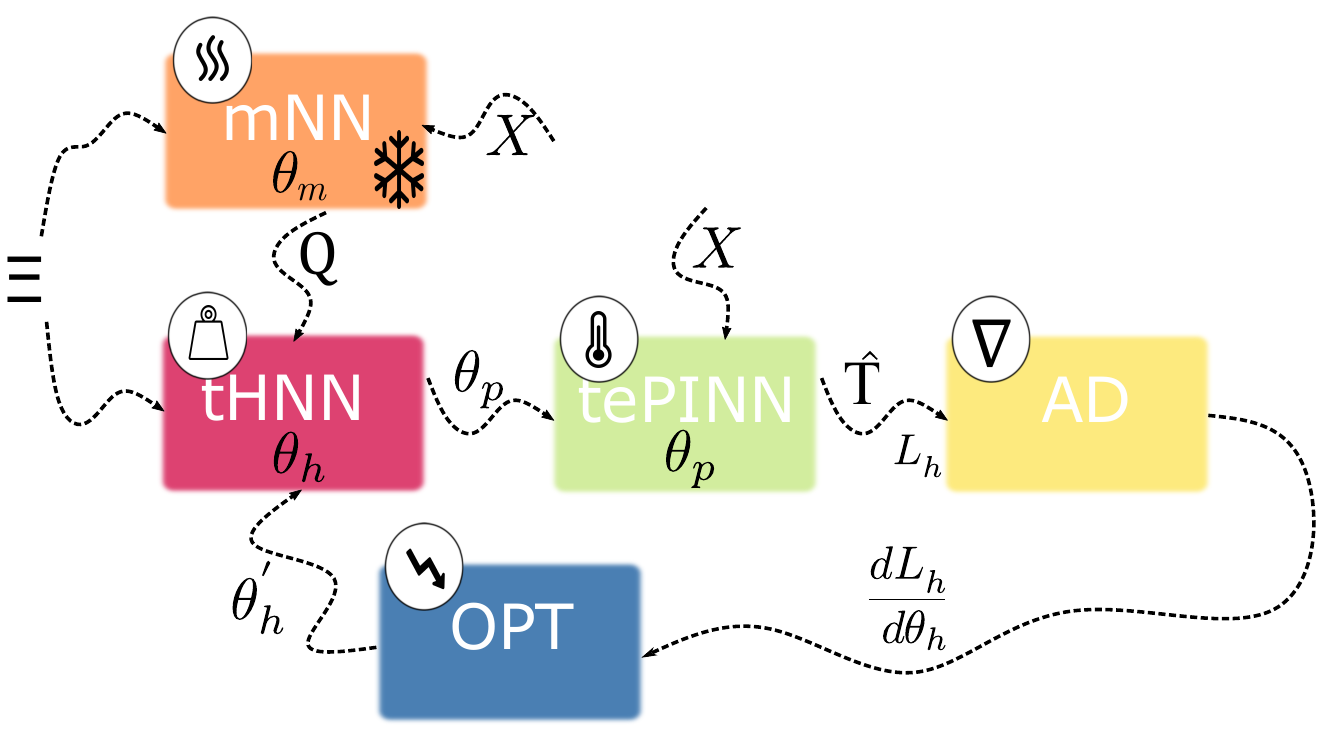}}
	\caption{Training scheme of tHNN involving a coupled magnetic-thermal problem. mNN has been previously trained (it has now frozen weights) and it outputs the heat source density at a given design parameter set $\pmb{\xi}$ evaluated at position X. tePINN receives the weights from tHNN and it predicts the temperature at location X. The objective to be minimized is $L_{h}$ (Eq. \ref{eq14}). \label{fig4}}
\end{figure}

Also tHNN is a fully-connected neural network. It has 4 inputs (the specific inductor geometry) and it outputs the weights and biases (it is indeed a hypernetwork) of a tePINN, now only having 2 hidden layers with 24 neurons each. Compared to the previous section, a "smaller" architecture (i.e., less tePINN weights) is preferred to limit the output dimension of the tHNN. Indeed, previous results show, that despite a sixfold reduction in the number of network weights (from 4417 to 697), the absolute maximum prediction error remains below 0.2$^\circ$C (Tab. \ref{tab05}). Since the output of tHNN are weights and biases of tePINN, limiting the dimensionality of the tHNN output could mean reducing the tePINN expressiveness. While we cannot completely rule out this issue, on the one hand, such a limitation does not occur in our work, on the other hand, hypernetworks have shown promising results in a variety of high-dimensional deep learning problems, including transfer learning, weight pruning, uncertainty quantification, zero-shot learning, natural language processing, and reinforcement learning \cite{r28}. 
tHNN has also 2 hidden layers, but the number of neurons is 128. ReLU is the chosen activation function. In the tHNN training (Fig. \ref{fig4}), for each system parameter set  $\pmb{\xi}$, mNN (already trained - it has now frozen weights and biases) predicts the heat source distribution $Q$ within the plate. In turn, the tHNN outputs the weights and biases $\boldsymbol{\theta_{p}}$ of a tePINN. tePINN provides the temperature field, evaluated at location $X$, for that parameter set. The temperature field and the heat source map are in fact required in the training loss (Eq. \ref{eq11}). This pass is repeated with all the inductor geometries in the batch ($N_{\xi}$ is 8) to obtain the tHNN loss function (Eq. \ref{eq13}). The tHNN is trained on 10$^{3}$ epochs with ADAM.

The temperature field provided by the trained tHNN is tested in case of 30 random inductor geometries. The MAE is 0.15$^{\circ}$C, while the maximum absolute error is 0.47$^{\circ}$C. This is almost a factor five higher than the single thermal analysis of previous section, but still represents an error lower than 0.5\% in relative terms. In fact, we are now dealing with a coupled problem in which both the magnetic and thermal solutions are approximated (in \ref{sec3c2}, the ground truth heat source distribution is considered, instead).

\subsection{The inverse problems: inductor optimal design}\label{sec3d}
\begin{figure}[!t]
	\centering{\includegraphics[width=0.5\columnwidth]{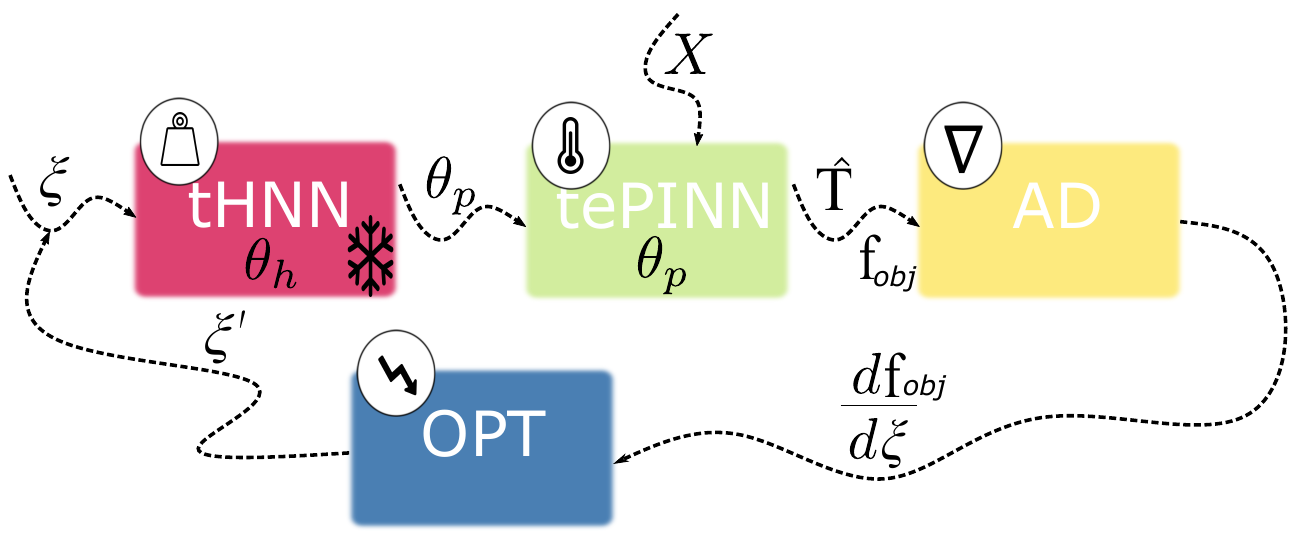}}
	\caption{Solving the inverse problem involving tHNN. tHNN has now frozen weights. tePINN receives the weights from tHNN and it predicts the temperature at location X. The objective to minimize is $f_{obj}$ and it comes from the design problem (Eq. \ref{eq18}). \label{fig5}}
\end{figure}

Once tHNN is trained, it is able to predict the entire temperature field for a given induction geometry. It takes two forward passes: in the first, the tHNN delivers the network weights of the corresponding tePINN, in the second, the tePINN predicts the temperature at locations $X$. It means that an inverse problem is generally fast to be solved. Moreover, due to AD, the optimizer could take advantage of the derivatives (Fig. \ref{fig5}). We note that the objective function(s) of the inverse problem are defined a-posteriori without posing any restriction to the training of the tHNN.

\subsubsection{Single-objective - uniform heating}
\label{subsingle}

In the following benchmark, the objective measures, using the "criterion of proximity", the temperature in-homogeneity. It is evaluated at the plate's upper boundary $\Gamma$ \cite{r18}:

\begin{equation}
	f_{prox}(\pmb{\xi}) = N_{\max} - N_{j\in \Gamma}(|T_j(\pmb{\xi})-T_{goal}|<\text{tol})
	\label{eq18}
\end{equation}

where N$_{\max}$ is the number of sampling points (200), $T_{goal}$ = 1130$^{\circ}$C is the temperature which should be reached in the plate, and tol = 10$^{\circ}$C (approx. 1\% in relative terms). We adopt a differential evolution (DE) as optimizer \cite{r22} and solve three subsequent inverse problems. Each is unconstrained and single-objective. That is, Eq. \ref{eq18} is minimized w.r.t. box constrained $\pmb{\xi}$. However, each method solves the field problem differently:

\begin{enumerate}
	\item[1)] tHNN: it involves tHNN for the parametric solution of the coupled problem (mNN is now no longer needed); 
	\item[2)] sNN: it relies on a "single neural network" (shortly sNN), trained to map the inductor geometry $\pmb{\xi}$ into a 200-dimensional vector that describes the temperature at the graphite plate's upper boundary. It uses 1296 (magnetic-thermal) FEA solutions as a data-set and it is trained minimizing the mean squared error between data and network prediction (we refer e.g. to \cite{r26} for more details). sNN is built on a standard supervised scheme;
	\item[3)] FEA: it relies only on FEA.  
\end{enumerate}

\begin{figure}[!t]
	\centering{\includegraphics[width=0.6\columnwidth]{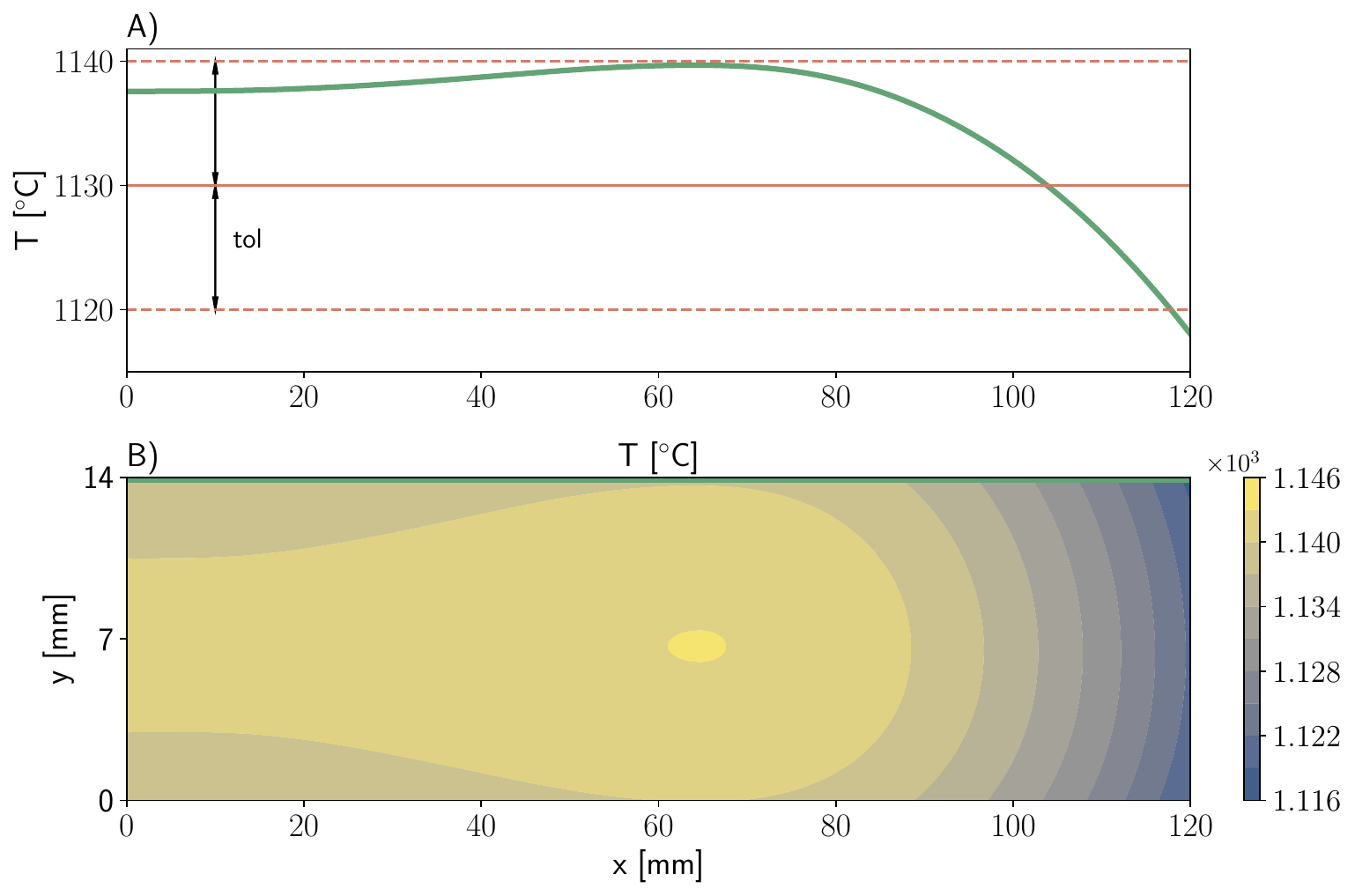}}
	\caption{Results of the inductor optimal design. \label{fig6} 
		A) Optimized temperature profile of the graphite plate upper bound. B) Optimized temperature field.}
\end{figure}

DE does not make use of gradient information that would be available with the tHNN and sNN due to automatic differentiation (in general, FEA performed with a commercial software does not supply derivatives). However, since the objective (Eq. \ref{eq18}) is based on a point counting function that works on discrete values, AD does not offer gradients in this case. All inverse problems achieve the same optimal objective of 2, showing similar accuracy and meaning that actually almost the entire upper graphite plate bound lies within the prescribed temperature tolerance (Fig. \ref{fig6}). Looking at the design variables, the discrepancy never overcomes 0.1 mm and 0.2 mm, for the tHNN and sNN, respectively (Tab. \ref{tab1}). 

\begin{table}
	\caption{Design variable and their ranges}
	\centering
	\begin{tabular}{lllllll}
		\toprule
		Design  & $l_b$  & $u_b$ & Start & Opt & Opt & Opt \\
		variable  & & & & tHNN & sNN & FEA \\
		\midrule
		$\xi_1$  & 5  & 15 & 14.8 & 5.1  & 5.0  &  5.1\\
		$\xi_2$  & 5  & 15 & 15  & 5.4  & 5.1 & 5.3\\
		$\xi_3$  & 5  & 15 & 15 & 10.9 & 11.1 & 11.0 \\
		$\xi_4$  & 5  & 15 & 15 & 14.9 & 15.0 & 14.9\\
		\bottomrule
	\end{tabular}
	\label{tab1}
\end{table}

The following remarks could be put forward. Solving the inverse problem with the third approach requires 1240 FEA magnetic-thermal weakly coupled solutions. Therefore, looking at the overall computational cost, tHNN and sNN are not computationally attractive. Indeed, tHNN employs 1296 magnetic solutions and it additionally requires training both the magnetic network mNN and the thermal hypernetwork tHNN. In turn, sNN builds on 1296 coupled FEA simulations and it requires the network to be trained beforehand. More in general, looking at the three approaches that solve the inductor design problem, the required steps could be classified as high, moderate, and low computationally intensive (Tab. \ref{tab2}). Each step is supposed to be performed either once or at each new (e.g., with a different objective) design problem. With no doubts, generating data (from FEA) to be provided to the sNN is  expensive. The same is valid for the tHNN because the magnetic network mNN is supervisely trained. We are currently working on solving also the time-harmonic problem in a fully unsupervised way, thus eliminating the need for FEA data. On the other hand, training both mNN and, in particular, tHNN is resource intensive (Tab. \ref{tab2}). Training the sNN is less demanding, since the network is usually smaller and it focuses on learning a specific quantity of interest. However, if the user intends to perform a design problem having a different objective, for instance, sNN must be retrained (or a new network is supposed to be trained). This is not the case for the tHNN. Finally, ultimate optimization (i.e., solving the design problem itself) is fast for both sNN and tHNN approaches. It takes around 110 seconds (DE population of 50 individuals, 50 generations) on a CPU of a personal laptop to solve the design problem with the tHNN. This value drops to 25 seconds with the sNN.

\begin{table}
	\caption{Qualitative computational cost of the required steps by different approaches (tHNN, single NN, FEA) when adopted in the solution of inverse design problems: \crule[red!90]{0.25cm}{0.25cm}(red) high-intensity, \crule[yellow!90]{0.25cm}{0.25cm}(yellow) moderate-intensity,\crule[green!90]{0.25cm}{0.25cm}(green) low-intensity. \label{tab2}}
	\centering
	\begin{tabular}{lllll}
		\toprule
		How often?  &  \multicolumn{2}{c}{Once} & \multicolumn{2}{c}{Each design} \\
		\midrule
		& Data Gen.  & Training & Training &  Optimization \\
		\midrule
		tHNN  & \crule[red!90]{0.25cm}{0.25cm}  & \crule[red!90]{0.25cm}{0.25cm}& -  & \crule[green!90]{0.25cm}{0.25cm} \\ \addlinespace[.35em]
		sNN  & \crule[red!90]{0.25cm}{0.25cm}  & - &  \crule[yellow!90]{0.25cm}{0.25cm}  & \crule[green!90]{0.25cm}{0.25cm} \\ \addlinespace[.35em]
		FEA  & - & - & - & \crule[red!90]{0.25cm}{0.25cm} \\
		\bottomrule
	\end{tabular}
\end{table}

Now is the critical point to consider when adopting the tHNN method. We need to determine the break-even point where it becomes advantageous compared to a fully FEA-based approach. Here are some cases:
\begin{enumerate}
	\item[A)] Whenever the user is undecided about the objective(s) (or constraints) of the design task, or he/she intends to perform multiple different design problems. This is especially true in the case of a multi-objective design (see section \ref{submco}); 
	\item[B)] If the user would like to get a fast solution of her/his design problem. Taking advantage of exact derivatives due to AD could be a plus;
	\item[C)] Whenever the design problem includes some hard constraints to be fulfilled. Usually, in the FEA context, since the derivative information is not available with commercial software, zero-order optimization methods are adopted. Possible constraints are therefore relaxed and they are treated as penalties in the objective function. In alternative, constraints could be converted into additional objectives, translating a constrained single-objective problem into an unconstrained multi-objective problem. Multi-objective problems will be addressed in section \ref{submco}. 
\end{enumerate}
Moreover, point B) is true also for the single network (sNN). Cases A) and C) would probably require additional training  (i.e., of other networks approximating the new objective(s) or constraint(s) - Tab. \ref{tab2}).

In order to demonstrate the acceleration of the optimal design solution when gradients are available (point B), we adopt, for a moment, a new objective function to minimize:
\begin{equation}
	f_{diff}(\pmb{\xi}) = \max_{j\in \Gamma} (|T_j(\pmb{\xi})-T_{goal}|)
	\label{eq19}
\end{equation}
where $T_j$ is temperature evaluated at N$_{\max}$ points located at the upper boundary of the plate $\Gamma$. Keeping the same box constraints for $\pmb{\xi}$ as before (Tab. \ref{tab1}), the sequential least squares quadratic programming (SLSQP) solver from Scipy \cite{r22} (linked with AD) solves the design problem involving tHNN in just 1 second. This is two order of magnitude faster than solving the previous problem (Eq. \ref{eq18}) with DE. The optimal design variables are [5.0, 5.0, 11.4, 15.0] and the optimal objective value is 10.6$^\circ$C.

To address the concern C), a constrained design problem is solved:
\begin{subequations}
	\begin{alignat}{2}
		&\!\min_{\pmb{\xi}\in[\pmb{\xi}_{\min}, \pmb{\xi}_{\max}]}        &\qquad& f_{diff}(\pmb{\xi}) 	\label{eq:optProb}\\
		&\text{subject to} &      & g(\pmb{\xi}) \leq 0			\label{eq:constraint1}
	\end{alignat}
	\label{eq20}
\end{subequations}
Taking Eq. \ref{eq19} as objective, to enforce a maximum temperature $T_{\max}$ throughout the whole plate out of the boundary ($\Omega \backslash \Gamma$), the constraint $g$ is:
\begin{equation}
	g(\pmb{\xi}) = \max_{j\in \Omega \backslash \Gamma} (|T_j(\pmb{\xi})|) - T_{\max}
	\label{eq21}
\end{equation}
$T_{\max}$ could be different from $T_{goal}$. Setting $T_{\max}$ at 1140$^{\circ}$C, the optimization problem (Eq. \ref{eq20}) using tHNN for the field analysis is solved with a SLSQP solver \cite{r22}. Again, gradient information is provided by AD for both objective and constraint. The solution is obtained within a 1 second. The optimal objective values 14.6$^\circ$C and the optimal position of the inductor turns are [5.0, 5.0, 10.3, 15.0]. In practice, in order to fulfill the temperature constraint, this is like the previous optimal inductor to which the third turn is moved downward by 1 mm.

\subsubsection{Multi-objective - uniform heating vs max power}
\label{submco}

Multi-objective optimization or Pareto optimization concerns with mathematical optimization problems involving more than one objective function to be simultaneously optimized simultaneously. Readers seeking additional information can refer to \cite{r27}. Two objectives are supposed to be minimized, namely the temperature uniformity along $\Gamma$ (Eq. \ref{eq18}) and the negative total induced power in the graphite plate (i.e., the induced power is maximized):
\begin{equation}
	f_{ind}(\pmb{\xi}) = -\int_{\Omega \ni X} Q(\pmb{\xi},X)d\Omega 
	\label{eq22}
\end{equation}

There are no other constraints besides box constraints. Again, the design problem is solved adopting three different methods for the coupled field anaylsis:
\begin{enumerate}
	\item[1)] tHNN: using tHNN for the thermal analysis and mNN for the magnetic one (current work); 
	\item[2)] sNN: employing a network trained on FEA data to predict the temperature field at boundary $\Gamma$ (just like in section \ref{subsingle}) and another to describe the heat map distribution (for convenience, mNN is taken);
	\item[3)] FEA: just relying on FEA simulations. 
\end{enumerate}

\begin{figure}[h!]
	\centering{\includegraphics[width=0.55\columnwidth]{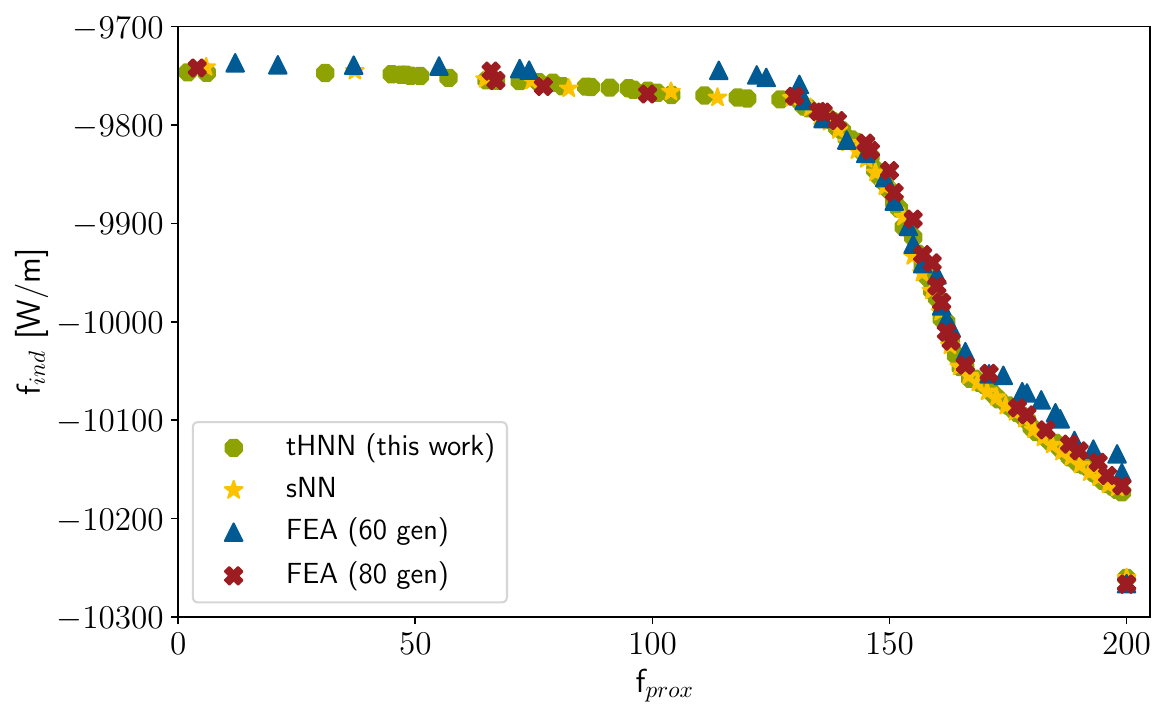}}
	\caption{Pareto frontiers with different approaches. \label{fig7}}
\end{figure}

NSGA-II is the chosen algorithm to solve the design task \cite{r29}. With a population of 50 individuals, 100 generations are performed with tHNN and sNN. To limit the computational cost, a population of 30 individuals is preferred with FEA. The problem is solved twice, by 60 and 80 generations, respectively. It turns out that tHNN, sNN, and FEA after 80 generations show comparable Pareto frontiers (Fig. \ref{fig7}), meaning that tHNN and sNN display a satisfactory accuracy. FEA requires 2430 field solutions. This is computationally more expensive than the 1296 magnetic simulations needed by the mNN (i.e., tHNN) training. sNN makes use of 1296 coupled solutions. Therefore, depending on the network training cost, sNN and tHNN could be computationally attractive. Since FEA is usually solved using CPUs, while network training uses GPUs, it is not straightforward to establish a fair criterion for measuring the total cost of different methods. The total time (for training, optimizing) is substantially influenced by the type of hardware used and it does not prove to be a valid criterion. Nevertheless, in the absence of a valid alternative, we can only say that sNN is as much computationally intensive as the FEA for the presented multi-objective problem. Conversely, the time taken to solve the problem using an FEA method is still lower than the cost of generating data for the mNN and training mNN and tHNN. However, tHNN becomes a time-saving alternative when the user would perform this multi-objective design after a single-objective design (section \ref{subsingle}). In fact, solving inverse problems with tHNN is almost inexpensive (Tab. \ref{tab2}). This, then, is the recommended application of this approach. It takes around 550 seconds to solve the multi-objective design (sNN needs 230 seconds instead). Please note that NSGA-II does not make use of derivative information. We are confident that approximating the Pareto frontier could be further speed up using gradient-based approaches.

\section{Conclusion}\label{sec4}
In this paper, we introduce the use of energy-based PINNs for solving the steady-state heat conduction equation. Compared to the residual-based approach, it shows multiple advantages, including a drastic reduction of the training time (even a factor 50 in the considered benchmark) and no use of hyperparameters in the loss function. On the contrary, it requires performing integration that, with the Gauss quadrature method, needs a sampling strategy close to a meshing step in conventional FEA. The parametric solution of the magnetic-thermal coupled problem makes use of FEA simulations to approximate the magnetic solution, while the thermal analysis results entirely unsupervised (due to the hypernetwork). The proposed framework achieves an accuracy that allows to solve an optimal inductor design problem (therefore involving a multi-physics analysis) with an objective that is defined a-posteriori (i.e., after the hypernetwork's training). Such a framework could be adopted, e.g., in the fast prototyping of inductors, due to both the freedom in defining the design objective(s) and little computational cost. A multi-criteria design is also straightforward. With reference to the analyzed case study, looking at the overall computational cost, the hypernetwork  becomes computationally attractive when a multi- and a single-objective design problems are solved. The obtained results are promising and motivate continuation of the researches and extension of them to the design of, e.g., various industrial electromagnetic devices. We are currently working on the solution of time-harmonic problems in a fully unsupervised way, thus eliminating the need for FEA data when setting up the hypernetwork for the thermal analysis.

\bibliographystyle{unsrtnat}

\end{document}